\documentclass{llncs}

\usepackage{times}
\usepackage[german,english]{babel}      
\usepackage[latin1]{inputenc} 
\usepackage[pdftex]{graphicx}       
\usepackage{array}          
\usepackage{amsmath}        
\usepackage{amssymb}        
\usepackage{amsfonts}        
\usepackage{mathrsfs}       
\usepackage{fancyhdr}
\usepackage{wrapfig}
\usepackage{url}
\usepackage{multirow}
\usepackage{rotating} 
\begin{document}
\title{Interest-Based vs. Social Person-Recommenders in Social
Networking Platforms}

\author{Georg Groh\inst{1} \and Michele Brocco\inst{1} \and Andreas Kleemann\inst{2}}
\institute{
Fakultät für Informatik,
Technische Universität München, Germany
\and
Utopia AG, München, Germany
}
\email{grohg,brocco,@in.tum.de; kleemu@web.de}

\maketitle
\begin{abstract}
Social network based approaches to person recommendations are compared to interest based approaches with the help of an empirical study on a large German social networking platform. We assess and compare the performance of different basic variants of the two approaches by precision / recall based performance with respect to reproducing known friendship relations and by an empirical questionnaire based study. In accordance to expectation, the results show that interest based person recommenders are able to produce more novel recommendations while performing less well with respect to friendship reproduction. With respect to the user's assessment of recommendation quality all approaches perform comparably well, while combined social-interest-based variants are slightly ahead in performance. The overall results qualify those combined approaches as a good compromise. 
\end{abstract}
\section{Introduction}
The term Social Recommender Systems can be understood in a variety of ways. The first interpretation of the term may substitute the actors of some sub-network of a social network for the set of users with similar rating-behavior as a neighborhood for making collaborative recommendations (see e.g. \cite{sinha2002}, \cite{bonhard2006} or \cite{grohEhmig}).

This approach has been shown to possess certain advantages over traditional collaborative filtering \cite{herlocker2002}
and has been shown to be able to perform as good or better at least in taste related domains \cite{grohEhmig}\cite{bonhard2006}. The advantages encompass a better performance in certain situations in view of the portfolio effect (a user is recommended items which he already knows or which are too similar to those he already knows (see \cite{billsus1998})) or cold start effects (with respect to ratings and trust \cite{golbeck2005}). New influences (structurally or radically new recommendations) can enter the information space of a user through sub-networks of her social network (horizon broadening effect) as easy of even easier than through groups of similar rating but otherwise unrelated users. But the external intelligence injected into the system by explicifying and using direct or indirect social relations as in social networking platforms may ensure a somewhat higher probability of relevance of those radically new and unexpected, horizon broadening recommendations for the user. Reasons for this can be normative effects in groups \cite{kaplan1987}\cite{neumann1991} (``I should know and like what my peer group likes'') or trust effects and easier explanations for recommendations via the social network (``I trust and know how to value this album recommendation based on Mark's, Jenny's and Yiming's musical information space because I know them and their relation to me and their function, role, position etc. in the network''). There may, however, be some cases where social recommenders in this first interpretation may be less useful than recommendations by ``network'' of anonymous but similar rating users (e.g. in case of recommendations of scientific papers) where implicit ``topical'' relations to these users are exploited.          

A second interpretation of the term Social Recommender System may encompass recommending items not to single users but to whole groups of users. In this interpretation the target of the recommendation (the group) is a socially defined concept \cite{woerndlBirnkammererGroh}. 

A third interpretation of the term may make persons or groups of persons the recommended entities, either using social filtering (as discussed above in the first interpretation of the term), conventional collaborative filtering, content-based filtering (using any accessible electronic representation of the recommendable persons or groups as a basis for similarity computations) etc.. One example for this interpretation are team recommendation systems (see e.g. \cite{brocco}) where teams are recommended (e.g. to HR administrators) especially in situations, where the number possible team configurations is very high (such as Open Innovation scenarios). 

This contribution will deal with a flavor of third interpretation: In a social networking platform recommending potentially interesting other users to a user based on mutual interests. While user recommendations on the basis of the social network are quite common today (consider e.g. the friend recommendations in Facebook), the problem of how to assess and incorporate user's interests into the recommendation in a simple and expressive way is still subject to research. In this article, we investigate the question, how simple interest based person recommendation approaches performs in contrast to social network based recommendation approaches.  

In Section \ref{rw} we review more related work concerning social recommender systems. Section \ref{dm} describes the setting of our study, the data-set we use, and the range of recommendation methods we investigate. Section \ref{rd} then presents and discusses the results. On the one hand we compute performance measures on the basis of reproduction of known friendships as indicators of the usefulness of the recommenders. Furthermore the results of an empirical study among our test-users is presented and discussed. In the conclusion, we summarize the overall results and shortly discuss and compare the implications for the various approaches to person recommendations.   
\section{Related Work} \label{rw}
Currently the most common example for people recommendation is people match-making in social networks. Popular application domains therein are dating platforms or expert finders. The first, may consider also preferences on different scales besides traditional demographical data such as age, gender, etc. The second considers skills, competencies and expertise acquired from various sources in order to recommend an expert that could provide suggestions and help for a specific problem.

In online dating platforms persons usually fill in data that should describe themselves appropriately in order to find partners that match their person in terms of demography (e.\,g. age, gender, height), interests (e.\,g. para-gliding, watching movies) and preferences (e.\,g. rock music, smoking). A few of them allows for entering a so called ``target profile'' that represents the description of the person one would like to receive as recommendation. Fiore et al. for instance, investigated which data within a profile influences the perceived attractiveness by women and men by correlating the perceived attractiveness with various elements of a user's profile (photos, free-text components and fixed-choice components) \cite{fiore2008} .

Diaz et al. \cite{diaz2010} describe the match-making problem from an information retrieval perspective and propose a novel approach for the combination of user profiles to improve the relevance of recommendations. There, features are extracted from a user profile (e.\,g. free-text descriptions) and used as input for a machine learning algorithm that selects the most important predictors for good matches. Good matches were considered those matches where bilateral user interaction could be identified or the same features applied as in the conditions where a bilateral contact occurred (cp. labeled vs. predicted relevance).

Expert finders are a different domain for person recommendations. There primarily competencies are regarded in order to increase the probability to find a solution for an occurred problem. In the simplest case, the task of finding experts can be solved with simple database queries. However, this does not always entail satisfactory results due to the difficulty to formulate appropriate queries and because skills may not be the only criterion for searching. McDonald and Ackerman \cite{mcdonald2000} for instance tried to model current best practices for finding experts in a large company and mapped these heuristics to a corresponding system.

In another work McDonald augmented this system with two different social networks: one based on workplace sociability, which represents how often individuals socialize with each other, the other based on shared workplace context, which represents logical work groups and work context over organizational boundaries \cite{mcdonald2003}. His work emphasized that it is challenging to mix skills and social networks in recommendations because users perceive a trade-off between the two: more precisely even though the system looks first for experts and ranks them afterwards according to the social network, the users think they get only recommended due to the latter aspect. A related system described by Ehrlich et al. \cite{ehrlich2007} was used (among other functionalities) to recommend experts searched by keywords within a specific social distance in a user's social network. Furthermore, this work addresses also other aspects such as privacy, acceptance and usage of such systems.

Guy et al. \cite{guy2008} describe a slightly different way of recommending persons. Their approach bases on the collection of data (from blogs, social bookmarking, etc.) in a company's internal intranet in order to make suggestions for adding people that may belong to its social network but were not explicitly added to the social networking platform.

Regarding person recommender system in enterprise-internal social networking platforms, Chen et al. \cite{chen} developed a person recommender based on keyword extraction algorithm, that tries to extrapolate user interest from user contributions. This approach is very valuable as foundation for our purpose, even if the findings of enterprise-internal platforms can not be inherently applied to comminties of interest such as the utopia community. Additionally, in contrast to Chen et al. we rather want to extrapolate user interests from all user activities performed in the predefined topic categories provided by the platform and hence investigate whether this kind of categorization technique is suitable as background data for interest-based person recommendations. As a last aspect, the recommendation provided by Chen et al.'s system, may have a different kind of utility or goal (see Section \ref{dm}) because of the business domain the community is situated in.

An extensive review on social matching systems is provided by Terveen and McDonald \cite{terveen2005} that additionally formulate claims and related research questions in this field. This work can be used to derive guidelines for the development of systems that incorporate social aspects (especially social networks) to find appropriate matches.
\section{Data-Set and Methods}  \label{dm}
We had access to the complete database from the German based social networking platform Utopia.de \cite{utopia}. The main purpose of the platform is the collaborative promotion, discussion and development of ideas and concepts contributing to more environmental sustainability. The platform provided the usual set of services and data-elements, like private messaging, discussion boards or blogs and personal profiles, which, in contrast to platforms targeted at self presentation, are rather sparse and use few pre-determined elements. In contrast to social network based friend recommendations (which we will refer to in the following as friend of a friend (FoF) recommendations) being widespread in Social Networking platforms or content-based recommendations comparing only the profiles directly, we are interested in recommending users other users on the basis of their interests reflected by their actions and content on the platform. Thus the sparse profiles are not a problem. 

Besides user generated contributions, the platform also contains editorial material which can be commented upon by the users. Furthermore, users can express positive attitudes toward a contribution by assigning it a ``worth living'' point. 
Instead of a free social tagging system, the platform has eleven content categories $(C_1,\ldots,C_{11})$ like e.g. ``Health and Diet'', ``Construction and Renovation'' that users can attach to any contribution, which can be viewed as a simple form of tagging with a fixed tag set. Social tagging based person recommender systems (e.\,g. \cite{zan}\cite{naka}), in general, recommend persons according to similar tagging behavior. In contrast to classic Collaborative Filtering (CF) which basically uses similarity measures on the columns of the user-item-rating matrix $R_{\{ui\}}$ for neighborhood creation towards recommending items, these systems use the user-tag-item matrix $T_{\{uti\}}$ to identify users with similar tagging behavior for recommending these persons (e.g. as a means of expert finding). Classic CF belongs to a class of recommending approaches that use explicit ratings of items (for item recommendation) or of persons (for person recommendation), whereas  social tagging based approaches belong to a class of methods that use implicit methods. Implicit methods induce user attitude towards items or similarity to other users indirectly from their behavior on the platform (e.g. frequency of accessing certain contributions) or the content in their information spaces (their contributions on the platform or their profile, which can be compared using techniques from information retrieval (e.g. tf-idf vectors and cosine similarity)). 

For person recommendations, users with similar tagging behavior can be considered to have ``similar interests'' and are thus candidates for being recommended. In essence, this interpretation is not necessary. The term ``users with similar interests'' can be considered synonym for ``users that are similar with respect to their behavior on the platform and / or their information spaces''. 

In contrast to having to compare users by comparing matrices as $sim(u_1,u_2)=sim(T_{u_1\{ti\}},T_{u_2\{ti\}})$ as in social tagging based person recommenders, a simpler approach is to count all platform activities of a user related to a certain category $C_i$ (``categorized activities''). Such categorized activities can be the creation or commenting of a content item (e.g. a blog entry) or the assignment of a ``worth living'' point. For each user $u$, these counts $act_{i}$ are then normalized with the total number of categorized activities $\sum_i act_{i}$, to yield the
normalized categorized activities $A_i = act_{i} / \sum_i act_{i}$. We can then compare these vectors as 
$sim(u_1,u_2)=sim(A_{u_1\{i\}},A_{u_2\{i\}})$ to yield a similarity measure for users. For the actual comparison of the vectors, we use standard cosine similarity and Pearson correlation. 
A minimum total number of categorized activities is necessary to be included in the matrix. 
From the resulting similarity matrix, we recommend users with a similarity above an adjustable threshold value that were not already ``friends''. 

Unfortunately, in the absence of a free social tagging mechanism in the platform we cannot compare the performance of our approach against the social tagging based person recommenders discussed before. 

An interesting question regarding the validation of person recommender approaches is the performance metric to be used. At this point a diversification of the different recommender goals should be done since two possible goals can be generally pursued. One possible evaluation method is to evaluate whether the user accepted the recommendation (for instance by clicking on the recommended item). The other possible evaluation is whether the recommendation itself is useful, i.\,e. if the person recommended in fact does fulfill a user's expectations (with respect to a predefined goal such as e.\,g. for a friendship, as discussion partner, as expert). Obviously, a strict diversification of goals is not possible, because (i) the goals are conceptually not completely disjunct and (ii) from a technical point of view platforms do not provide this diversification for classifying users. Thus, friendships in social networks are treated as a sort of ``bookmarks'' to find persons with respect to all the above mentioned goals. In the utopia case, that can be regarded as a community of interest, we want people to get in contact aiming at finding new discussion partners such that interesting hints and suggestions related to the topics discussed in the utopia platform can be better exchanged.

However, as mentioned the target platform does not provide any diversification concerning this aspect. For this reason, and knowing that the goals for person recommendation in our case overlap, as a measure for the
potential success of the approach we had to choose the reproduction rate of friendship ties that already exist in the platform as evaluation criterion.
Therefore, and also in order to compare the approach against a FoF based approach, we exclude members with less than 3 friends and less than 8 friends of friends within the test group. The minimum total number of categorized activities was set to 3. The resulting test group encompassed 334 users with 3984 friendship relations. The mean number of friends was $11.93$. mean number of FoF was $270.31$ and mean number of categorized activities was $87.56$. $\sim$31\% of the test users had only 3 or 4 friends, and only $\sim$17\% of the users had only three or four categorized activities. 

The FoF Recommender that is similar to the friend recommenders used in common Social networking platforms (see Section \ref{rw}) and that we compare our approach against, recommends a person $u_1$ to a person $u_2$ in proportion to the number of common friends $f_{u_1\wedge u_2}$ relative to the average total number of friends of both users $0.5(f_{u_1}+f_{u_2})$. The FoF similarity or recommendability of $u_1$ and $u_2$ is thus given by $sim(u_1,u_2)= 2 f_{u_1\wedge u_2} / (f_{u_1}+f_{u_2})$.

We used 10-fold cross validation in our experiments: For each of 10 runs of all of the recommender approaches that we compare, we leave out one tenth of the friendship relations and use the remaining nine tenth of the friendship relations to compute FoF Recommendations. The data basis for most of the variations of the basic interest based recommendation approach, which will be discussed in the next section, remains constant. 

We then compute the $n=10$ best recommendations for each user and each approach and measure how many of the deleted one tenth friendship relations are ``reproduced'' by the recommender. 
If we recommend a total of $A$ persons in one run and have 398 deleted friends per run, we can determine the true and false  positives ($TP$ and $FP$) and the false negatives $FN$ and have $A = TP + FP$ and $FN= 398 - TP$. We can then compute Precision, Recall and F-Measure as usual as measures of the success rate. 

If the random 10-fold partitioning of the friendship relations deletes less than 1 or more than 10 friendship relations for a single user, we do not compute recommendations for this user. In these cases we cannot determine the success rate analogous to the ``regular'' cases. Thus from the 334 (users) $*$ 10 (runs) $=$ 3340 cases we only compute recommendations for 1921 of these cases. Since for each case we recommend the top $n=10$ best recommendations we make 19210 recommendations in total. If we recommend a person that is already a friend in the respective nine tenth relation data set, we drop this recommendation. 

With this procedure, we can, of course never reach precision values of 1, simply because we delete on average in each run only $2.07$ friend relations and recommend almost always 10 persons. However, these restrictions apply to all recommenders equally. 
\section{Results and Discussion} 
\label{rd}
A general strength of the proposed approach can be that, in contrast to social network based approaches like FoF, a user does not need a friend-list, but a weak point is that passive users that do not perform many explicit actions will not acquire a meaningful $A_{\{i\}}$ vector.  
\begin{table}[htb]
\begin{center}
\begin{tabular}{|l||l|l||l|l|l|l|l|}
\hline
\footnotesize Recommender 
&\begin{sideways} \parbox[t]{17ex}{\footnotesize total $\#$ recommen-\\dations} \end{sideways}
&\begin{sideways} \parbox[t]{17ex}{\footnotesize true positives (TP) \\ (reproductions)} \end{sideways}
&\footnotesize\begin{sideways}\footnotesize precision\end{sideways}
&\footnotesize\begin{sideways}\footnotesize recall\end{sideways} 
&\footnotesize\begin{sideways}\footnotesize f-measure\end{sideways}\\ \hline\hline
\footnotesize Random   & \footnotesize 19210 & \footnotesize 144 & \footnotesize 0.008 & \footnotesize 0.036 & \footnotesize 0.012 \\ \hline
\footnotesize Interest Based Pearson   & \footnotesize 19048 & \footnotesize 250 & \footnotesize 0.013 & \footnotesize 0.063 & \footnotesize 0.022 \\ \hline
Interest Based  Cosine  & \footnotesize 19210 & \footnotesize 283 & \footnotesize 0.015 & \footnotesize 0.072 & \footnotesize 0.024 \\ \hline
\footnotesize FoF   & \footnotesize 19132 & \footnotesize 1164 & \footnotesize 0.061 & \footnotesize 0.294 & \footnotesize 0.101 \\ 
\hline\hline
\footnotesize Interest Based Pearson plus link   & \footnotesize 19048 & \footnotesize 376 & \footnotesize 0.020 & \footnotesize 0.095 & \footnotesize 0.033 \\ \hline
Interest Based Cosine plus link   & \footnotesize 19210 & \footnotesize 422 & \footnotesize 0.022 & \footnotesize 0.107 & \footnotesize 0.036 \\ \hline
\end{tabular}
\end{center}
\caption{Results of the experiment. 10-fold cross validation: precision, recall and f-measure averaged over 10 runs.}
\label{crossvaltable1}
\end{table}

Table \ref{crossvaltable1} shows the basic results of the experiment. What we see from the table is that the interest based recommender approach is significantly better than random in reproducing pre-existing friendships. The FoF approach is even significantly better. This can be attributed to the fact that even in a platform that is mainly targeted towards exchange of content in view of a narrower field of interest (a typical community of interest \cite{carotenuto}), friendship relations are perceived mainly as something social and not so much as something content or interest related. 
The formation of social friendship ties will obviously be strongly influenced by the friend of a friend effect and can thus much more easily be reproduced by the FoF recommender. However, our approach does not aim at recommending friends in the mere social sense, but rather at recommending persons that are related via interests in platforms where the exchange in terms of content is the main goal as opposed to platforms where self-presentation and socially related communication is predominant. Of course, the social sphere and the interest based sphere are closely related. 

An attempt to nevertheless improve our interest based approach, we investigated, if weighting different sorts of categorized actions differently can make a difference (e.g. by giving creating a long blog-entry a higher weight than just assigning some item a ``worth living'' point). The variations only very slightly improved the performance (e.g. in the Pearson case plus 4$\%$ for precision), which does not allow for any significant conclusions. 

In \cite{chen}, authors were able to improve a content based item recommender system by additionally taking into account social relations between the item's owners. In accordance to that we also investigated, in how far our interest based person recommendation approach may profit from combining it with a social relation based component. We do this by multiplying the relevant ($\ge 0.5$) interest based similarity scores between two users with a factor of $1.5$ if the two users have at least one common friend. 
Thus we effectively augment the interest based approach with the FoF approach (and not vice versa). Thus the general advantages of our approach discussed before can be maintained. We called these variations ``plus link'' and the results are also shown in table \ref{crossvaltable1}. As expected, we see that the approach can profit from this augmentation by approximately 50 \% increase of performance. However, it has to be stressed that, as discussed before, the performance with respect to reproducing existing friendship ties is certainly not our main goal and not the only quality criterion of an interest based person recommender in our sense. 
\subsection{Empirical study}
\begin{table}[htb]
\begin{center}
\begin{tabular}{|l|l|l|}
\hline 
\footnotesize Q-nr & \footnotesize Question & \footnotesize Scale \\ \hline \hline
\footnotesize 1 & Do You know this user already? & \footnotesize yes / no \\ \hline
\footnotesize 2 & Are You interested in getting to know this user? & \footnotesize 5 point Lickert \\ \hline
\footnotesize 3 & Space for comments on the recommendations & \footnotesize Free text field \\ \hline
\footnotesize 4 & Are You generally interested in getting to know new users on Utopia? & \footnotesize 5 point Lickert  \\ \hline
\footnotesize 5 & Would You like to be recommended users on Utopia? & \footnotesize 5 point Lickert \\ \hline
\footnotesize 6 & Space for general comments & \footnotesize Free text field \\ \hline
\end{tabular}
\end{center}
\caption{Online survey: questions}
\label{questions}
\end{table}

In order to address this issue, we conducted an online empirical study among our test users, dividing them into three groups and providing them with 5 person recommendations using one sort of recommender in each group (Interest based cosine, interest based cosine plus link and FoF). The users were asked to evaluate the recommendations according to several criteria (see table \ref{questions}). General statistics with respect to this survey are shown in table \ref{statistics}. 

%
\begin{table}[htb]
\begin{center}
\begin{tabular}{|p{0.3\columnwidth}||p{0.2\columnwidth}|p{0.2\columnwidth}|}
\hline 
\footnotesize Recommender & \footnotesize Completed questionnaires & \footnotesize Percent completed \\ \hline \hline 
\footnotesize Cosine & 28 & 28.3 \% \\ \hline
\footnotesize Cosine plus link & 35 & 35.4 \% \\ \hline
\footnotesize FoF & 36 & 36.4 \% \\ \hline
\end{tabular}
\end{center}
\caption{Online survey: general statistics}
\label{statistics}
\end{table}
The recommendations were provided in the form of picture and username of the recommended person as specified in the platform. By clicking on either username or picture, the profile page of the corresponding person could be inspected in order to identify possible interesting characteristics of that person. Based on this knowledge the user can decide whether the recommendation proposed is appropriate or not.

\begin{table}[htb]
\begin{center}
\begin{tabular}{|p{0.2\columnwidth}||p{0.19\columnwidth}|p{0.19\columnwidth}|p{0.07\columnwidth}|}
\hline
 \footnotesize Recommender & \footnotesize rec. person known & \footnotesize rec. person unknown & \footnotesize \# of rec. \\ \hline\hline
 \footnotesize Cosine & \footnotesize 40,7\% (57) & \footnotesize 59,3\% (83) & \footnotesize 140 \\ \hline
 \footnotesize Cos.~plus Link & \footnotesize 41,1\% (72) & \footnotesize 58,9\% (103) & \footnotesize 175 \\ \hline
 \footnotesize FoF & \footnotesize 57,8\% (104) & \footnotesize 42,2\% (76) & \footnotesize 180 \\ \hline \hline
 \footnotesize Overall & \footnotesize 47,1\% (233) & \footnotesize 52,9\% (262) & \footnotesize 495 \\ \hline
\end{tabular}
\end{center}
\caption{Results of question 1}
\label{surveynoveltytable}
\end{table}
Table \ref{surveynoveltytable} shows that the FoF variant is more likely to recommend already known persons which is socially plausible. The overall high number of recommendations of already familiar persons can be explained by the fact that due to the selection scheme of the 334 users (see previous section), already very active users were selected that have a high probability of knowing each other. 

\begin{table}[htb]
\begin{center}
\begin{tabular}{|p{0.1\columnwidth}||p{0.15\columnwidth}|l|l|l|l|l|}
\hline
\multirow{2}{*}{\footnotesize Rec.} & \multirow{2}{*}{\footnotesize Quest.~1} & \multicolumn{5}{c|}{\footnotesize Question 2}\\ \cline{3-7}
& & \footnotesize 1 & \footnotesize 2 & \footnotesize 3 & \footnotesize 4 & \footnotesize 5 \\ \hline\hline
\multirow{3}{*}{\footnotesize Cos.} & \footnotesize (unknown) & \footnotesize 24.1\% & \footnotesize 16.9\% & \footnotesize 34.9\% & \footnotesize 16.9\% & \footnotesize 7.2\% \\ \cline{2-7}
 & \footnotesize  (known) & \footnotesize 1.8\% & \footnotesize 3.5\% & \footnotesize 54.4\% & \footnotesize 26.3\% & \footnotesize 14.0\% \\ 
\cline{2-7}
 & \footnotesize  Overall & \footnotesize 15.0\% & \footnotesize 11.4\% & \footnotesize 42.9\% & \footnotesize 20.7\% & \footnotesize 10.0\% \\ 
\hline
\multirow{3}{*}{\footnotesize \parbox{0.1\columnwidth}{Cos. \\pl.~lnk.}} & \footnotesize (unknown) & \footnotesize 5.8\% & \footnotesize 27.2\% & \footnotesize 28.2\% & \footnotesize 27.2\% & \footnotesize 11.7\% \\
\cline{2-7}
 & \footnotesize  (known) & \footnotesize 22.2\% & \footnotesize 6.9\% & \footnotesize 23.6\% & \footnotesize 26.4\% & \footnotesize 20.8\% \\
\cline{2-7}
 & \footnotesize  Overall & \footnotesize 12.6\% & \footnotesize 18.9\% & \footnotesize 26.3\% & \footnotesize 26.9\% & \footnotesize 15.4\% \\   
 \hline
\multirow{3}{*}{\footnotesize FoF} & \footnotesize (unknown) & \footnotesize 9.2\% & \footnotesize 19.7\% & \footnotesize 40.8\% & \footnotesize 14.5\% & \footnotesize 15.8\% \\ \cline{2-7}
 & \footnotesize  (known) & \footnotesize 15.4\% & \footnotesize 1.9\% & \footnotesize 33.7\% & \footnotesize 26.9\% & \footnotesize 22.1\% \\ 
\cline{2-7}
 & \footnotesize  Overall & \footnotesize 12.8\% & \footnotesize 9.4\% & \footnotesize 36.7\% & \footnotesize 21.7\% & \footnotesize 19.4\% \\ 
 \hline\hline
\multirow{3}{*}{\footnotesize Overall} & \footnotesize (unknown) & \footnotesize 12.6\% & \footnotesize 21.8\% & \footnotesize 34.0\% & \footnotesize 20.2\% & \footnotesize 11.5\% \\ 
\cline{2-7}
 & \footnotesize  (known) & \footnotesize 14.2\% & \footnotesize 3.9\% & \footnotesize 35.6\% & \footnotesize 26.6\% & \footnotesize 19.7\% \\ 
\cline{2-7}
 & \footnotesize  Overall & \footnotesize 13.3\% & \footnotesize 13.3\% & \footnotesize 34.7\% & \footnotesize 23.2\% & \footnotesize 15.4\% \\ 
 \hline
\end{tabular}
\end{center}
\caption{Cross-table question 1 (Familiarity) and question 2 (Interestingness)}
\label{tab:surveyratingcrossknown}
\end{table}
Table \ref{tab:surveyratingcrossknown} shows the relation between previous familiarity and the rating of interestingness. We see that the error of central tendency is present throughout the results of question 2. It is overall slightly more present for the recommender that does not make use of the social network (cosine). However, for the recommenders that make use of the social network (FoF and cosine plus link) this tendency is slightly more prominent for the unfamiliar recommended persons than for the familiar, while for the recommender that is purely interest bases (cosine) this slight effect is reversed. As an explanation, knowing a person may make it easier to come to an expressive estimation apart from the less meaningful middle rating. However, it also has to be taken into account that a main value for a recommender is to recommend new entities (persons in our case), where the use of these novel recommendations often can only be properly assessed a posteriori.    

\begin{table}[htb]
\begin{center}
\begin{tabular}{|l||r|r|r|r|r|}
\hline
\footnotesize Recommender & \footnotesize 1 & \footnotesize 2 & \footnotesize 3 & \footnotesize 4 & \footnotesize 5 \\ \hline\hline
\footnotesize Cos. & \footnotesize 21.4\% & \footnotesize 3.6\% & \footnotesize 35.7\% & \footnotesize 21.4\% & \footnotesize 17.9\%  \\ \hline
\footnotesize Cos.~pl.~lnk. & \footnotesize 20.0\% & \footnotesize 2.9\% & \footnotesize 17.1\% & \footnotesize 45.7\% & \footnotesize 14.3\%  \\ \hline 
\footnotesize FoF & \footnotesize 11.1\% & \footnotesize 11.1\% & \footnotesize 16.7\% & \footnotesize 41.7\% & \footnotesize 19.4\% \\  \hline \hline
\footnotesize Overall & \footnotesize 17.2\% & \footnotesize 6.1\% & \footnotesize 22.2\% & \footnotesize 37.4\% & \footnotesize 17.2\% \\  \hline
\end{tabular}
\end{center}
\caption{Results of Question 5 (General interest in person recommendation service)}
\label{tab:surveyglobalusetool}
\end{table}

\begin{table}[htb]
\begin{center}
\begin{tabular}{|c||r|r|r|r|r|}
\hline
\footnotesize Recommender & \footnotesize 1 & \footnotesize 2 & \footnotesize 3 & \footnotesize 4 & \footnotesize 5 \\ \hline\hline
\footnotesize Cosine & \footnotesize 0.0\% & \footnotesize 3.6\% & \footnotesize 32.1\% & \footnotesize 35.7\% & \footnotesize 28.6\% \\ 
\footnotesize Cos.-plus-Link & \footnotesize 2.9\% & \footnotesize 8.6\% & \footnotesize 28.6\% & \footnotesize 31.4\% & \footnotesize 28.6\% \\ 
\footnotesize FoF & \footnotesize 2.8\% & \footnotesize 2.8\% & \footnotesize 27.8\% & \footnotesize 30.6\% & \footnotesize 36.1\%  \\ 
\hline\hline
\footnotesize Overall & \footnotesize 2.0\% & \footnotesize 5.1\% & \footnotesize 29.3\% & \footnotesize 32.3\% & \footnotesize 31.3\%  \\ 
\hline
\end{tabular}
\end{center}
\caption{Results of Question 4 (General interest in getting to know new people)}
\label{tab:surveygloballikenetworking}
\end{table}

The results of the general questions of the questionnaire are shown in tables  \ref{tab:surveygloballikenetworking} and \ref{tab:surveyglobalusetool}. For question 4, we see that, according to expectation, the tendency to be interested in getting to know new people on a social networking platform is quite high. There are no significant differences among the three test-groups. With respect to question 5, we see that the recommendation service is regarded as overall positive but judged more critically (23.3 \% negative (rating 1 or 2) answers in question 5) compared to the general predisposition to be interested in getting to know new people (7.1 \% negative answers in question 4). However, the share of positive answers (rating 4 or 5) among the group which were confrontend with the recommendations from the merely interest based recommender (cosine) is significantly lower (39.3 \%) than for the groups that were confronted with recommendations that included the social network (60.0 \% and 61.1 \%). This can be seen as a hint that the social network plays an important role for people recommendations. However, the effect that the use of novel recommendations often can only be properly assessed a posteriori needs to be taken into account here as well, because according to table \ref{surveynoveltytable}, the cosine recommender proposes more novel recommendations than the FoF recommender.  The cosine plus link recommender that results in roughly the same share of novel recommendations as the cosine recommender appears to be a good compromise in view of this phenomenon. 
\section{Conclusion}
From our study it can be concluded that in social networking platforms, person recommenders are services that have some potential to deliver an added value for a large number of users. Purely interest based recommenders may produce more novel recommendations than purely social network based recommenders. With respect to the survey rating of test-users, the purely interest based approaches perform slightly worse than the purely social network based approaches. The over-proportionally good performance of the FoF approach in reproducing known friendships can be attributed to social effects and does not have to be taken as a definitive quality criterion. Mixed approaches yield many novel recommendation while (with respect to user rating) perform as good or even slightly better than the purely social network based approach. Combined social network based and interest based approaches may thus be a good compromise and a promising field of future research.   
\bibliographystyle{splncs03}
\bibliography{georgBibFileAbgekuerzt}

\begin{thebibliography}{10}
\providecommand{\url}[1]{\texttt{#1}}
\providecommand{\urlprefix}{URL }

\bibitem{utopia}
Utopia. \url{http://www.utopia.de}, (URL, Oct 2010)

\bibitem{billsus1998}
Billsus, D., Pazzani, M.J.: Learning collaborative information filters. In:
  Proc. 15th Int'l Conf. on Machine Learning. pp. 46--54 (1998)

\bibitem{woerndlBirnkammererGroh}
Birnkammerer, S., Woerndl, W., Groh, G.: Recommending for groups in
  decentralized collaborative filtering. Tech.-Report TUMI0927, TU Muenchen
  (2009)

\bibitem{bonhard2006}
Bonhard, P., Sasse, M.A.: 'knowing me, knowing you' -- using profiles and
  social networking to improve recommender systems. BT Technology Journal
  24(3),  84--98 (2006)

\bibitem{brocco}
Brocco, M., Groh, G.: A meta model for team recommendations in open innovation
  networks. Proc. ACM RecSys 09  (2009)

\bibitem{kaplan1987}
Caplan, M.F., Miller, C.E.: Group decision making and normative versus
  informational influence effects of type of issue and assigned decision rule.
  Pers. Soc. Psychol. Bull pp. 306--313 (1987)

\bibitem{carotenuto}
Carotenuto, L., et~al.: Communityspace: Toward flexible support for voluntary
  knowledge communities. Proc. Workshop on Workspace Models for Collaboration,
  London, UK, April 1999  (1999)

\bibitem{chen}
Chen, J., Geyer, W., Dugan, C., Muller, M., Guy, I.: Make new friends, but keep
  the old - recommending people on social networking sites. Proc. Int'l Conf on
  Human Factors in Computing Systems  (2009)

\bibitem{diaz2010}
Diaz, F., Metzler, D., Amer-Yahia, S.: Relevance and ranking in online dating
  systems. In: Proc. SIGIR10. pp. 66--73. ACM, New York, NY, USA (2010)

\bibitem{ehrlich2007}
Ehrlich, K., Lin, C.Y., Griffiths-Fisher, V.: Searching for experts in the
  enterprise: combining text and social network analysis. In: Proc. GROUP07.
  pp. 117--126 (2007)

\bibitem{fiore2008}
Fiore, A.T., Taylor, L.S., Mendelsohn, G., Hearst, M.: Assessing attractiveness
  in online dating profiles. In: CHI '08: Proceeding of the twenty-sixth annual
  SIGCHI conference on Human factors in computing systems. pp. 797--806. ACM,
  New York, NY, USA (2008)

\bibitem{golbeck2005}
Golbeck, J.A.: Computing and applying trust in web-based social networks. Ph.D.
  thesis, College Park, MD, USA (2005), chair-James Hendler

\bibitem{grohEhmig}
Groh, G., Ehmig, C.: {Recommendations in Taste Related Domains: Collaborative
  Filtering vs. Social Filtering}. Proc. Group07, Sunibel Island, USA, Nov 2007
   (2007)

\bibitem{guy2008}
Guy, I., Jacovi, M., Shahar, E., Meshulam, N., Soroka, V., Farrell, S.:
  Harvesting with sonar: the value of aggregating social network information.
  In: Proc. CHI08. pp. 1017--1026 (2008)

\bibitem{herlocker2002}
Herlocker, J.L., Konstan, J.A., Riedl, J.T.: An empirical analysis of design
  choices in neighborhood-based collaborative filtering systems. Information
  Retrieval  5,  287--310 (2002)

\bibitem{mcdonald2003}
McDonald, D.W.: Recommending collaboration with social networks: a comparative
  evaluation. In: Proc CHI03. pp. 593--600 (2003)

\bibitem{mcdonald2000}
McDonald, D.W., Ackerman, M.S.: Expertise recommender: a flexible
  recommendation system and architecture. In: Proc. CSCW00. pp. 231--240 (2000)

\bibitem{naka}
Nakamoto, R., Nakajima, S., Miyazaki, J., Uemura, S.: Tag-based contextual
  collaborative filtering. IAENG International Journal of Computer Science
  (2008)

\bibitem{neumann1991}
Noelle-Neumann, E.: The theory of public opinion: the concept of the spiral of
  silence. J. A. Anderson (Ed.): Communication Yearbook 14; Newbury Park, CA:
  Sage pp. 256--287 (1991)

\bibitem{sinha2002}
Sinha, R., Swearingen, K.: Comparing recommendations made by online systems and
  friends. In Delos-NSF Workshop on "Personalisation and Recommender Systems in
  Digital Libraries"  (2001),
  \url{http://www.rashmisinha.com/articles/Recommenders_Delos01.PDF (URL Jun
  2007)}

\bibitem{terveen2005}
Terveen, L., McDonald, D.W.: Social matching: A framework and research agenda.
  ACM Trans. Comput.-Hum. Interact.  12(3),  401--434 (2005)

\bibitem{zan}
Zanardi, V., Capra, L.: Social ranking: Uncovering relevant content using
  tag-based recommender systems. Proc. ACM RecSys 2008  (2008)

\end{thebibliography}
\end{document}